\documentclass[11pt,twoside]{article}

\usepackage{asp2014}

\aspSuppressVolSlug
\resetcounters

\begin{document}

\title{RCSEDv2: Processing and analysis of 4+ million galaxy spectra}

\author{Vladimir Goradzhanov,$^{1,2}$ Igor Chilingarian,$^{3,1}$ Evgenii Rubtsov,$^{1,2}$ Ivan Katkov,$^{4,5,1}$ Kirill Grishin,$^{6,1}$ Victoria Toptun,$^{1,2}$ Anastasia Kasparova,$^1$ Vladislav Klochkov$^{1,2}$ and Sviatoslav Borisov$^{7,1}$}
\affil{$^1$Sternberg Astronomical Institute, Moscow State University,  Moscow, Russia}
\affil{$^2$Faculty of Physics, Moscow State University,  Moscow, Russia}
\affil{$^3$Harvard-Smithsonian Center for Astrophysics,  Cambridge, USA}
\affil{$^4$New York University Abu Dhabi, UAE}
\affil{$^5$Center for Astro, Particle, and Planetary Physics, NYU AD, UAE}
\affil{$^6$Universit\'e de Paris, CNRS, Astroparticule et Cosmologie, F-75013 Paris, France}
\affil{$^7$Department of Astronomy, University of Geneva, Switzerland}

% This section is for ADS Processing.  There must be one line per author. paperauthor has 9 arguments.
\paperauthor{Vladimir Goradzhanov}{goradzhanov.vs17@physics.msu.ru}{0000-0002-2550-2520}{Sternberg Astronomical Institute, Lomonosov Moscow State University}{}{Moscow}{}{119234}{Russia}
\paperauthor{Igor Chilingarian}{igor.chilingarian@cfa.harvard.edu}{0000-0002-7924-3253}{Center for Astrophysics - Harvard and Smithsonian}{}{Cambridge}{}{02138}{USA}
\paperauthor{Evgenii Rubtsov}{rubtsov602@gmail.com}{0000-0001-8427-0240}{Sternberg Astronomical Institute, Lomonosov Moscow State University}{}{Moscow}{}{119234}{Russia}
\paperauthor{Ivan Katkov}{katkov.ivan@gmail.com}{0000-0002-6425-6879}{NYU Abu Dhabi}{Center for Astro, Particle, and Planetary Physics}{Abu Dhabi}{}{129188}{UAE}
\paperauthor{Kirill Grishin}{kirillg6@gmail.com}{0000-0003-3255-7340}{Sternberg Astronomical Institute, Lomonosov Moscow State University}{}{Moscow}{}{119234}{Russia}
\paperauthor{Victoria Toptun}{victoria.toptun@voxastro.org}{0000-0003-3599-3877}{Sternberg Astronomical Institute, Lomonosov Moscow State University}{}{Moscow}{}{119234}{Russia}
\paperauthor{Anastasia Kasparova}{anastasya.kasparova@gmail.com}{0000-0002-1091-5146}{Sternberg Astronomical Institute, Lomonosov Moscow State University}{}{Moscow}{}{119234}{Russia}
\paperauthor{Vladislav Klochkov}{vladislavk4481@gmail.com}{0000-0003-3095-8933}{M.V. Lomonosov Moscow State University}{Department of Physics}{Moscow}{}{119991}{Russia}
\paperauthor{Sviatoslav~Borisov}{sviatoslav.borisov@unige.ch}{0000-0002-2516-9000}{University of Geneva}{Department of Astronomy}{Versoix}{Canton of Geneve}{1290}{Switzerland}

\begin{abstract}
RCSEDv2 (https://rcsed2.voxastro.org/), the second Reference Catalog of Spectral Energy Distributions of galaxies, provides the largest homogeneously analyzed collection of optical galaxy spectra originating from several ground-based surveys collected between 1994 and 2019. The database contains astrophysical parameters obtained using the same data analysis approach from a sample of over 4 million optical spectra of galaxies and quasars: kinematics of stellar populations and ionized gas, chemical composition and age of stellar populations, gas phase metallicity. The dataset is available via Virtual Observatory access interfaces (IVOA TAP and SSAP) and through the web-site. Here we describe the RCSEDv2 spectroscopic dataset and the data processing and analysis.
\end{abstract}

\section{Spectroscopic data in RCSEDv2}

RCSEDv2 is a second generation of the RCSED project \citep{RCSED}, which included spectra and UV-to-NIR photometry for a sample of 800k galaxy spectra originating from SDSS DR7 \citep{sdssdr7}. In RCSEDv2 we have re-processed and analyzed over 4 million spectra from SDSS/eBOSS \citep{2020MNRAS.498.2354R}, Hectospec public archive \citep{2005PASP..117.1411F}, LAMOST \citep{2019ApJS..240....6Y}, 2dFGRS \citep{2001MNRAS.328.1039C}, WiggleZ \citep{2012PhRvD..86j3518P}, DEEP2/DEEP3 \citep{2013ApJS..208....5N}, FAST public archive \citep{2021AJ....161....3M}, and 6dFGS \citep{2004MNRAS.355..747J} in the redshift range ($0<z<1$). RCSEDv2 sample covers almost the entire sky and it has a substantially better completeness in the low-redshift end thanks to the magnitude limited redshift surveys UZC and 6dFGRS, and also a sample of galaxy spectra from LAMOST, which fill gaps in the SDSS/eBOSS redshift coverage at the low end. In addition to survey data, we included all public extragalactic spectra collected with Hectospec instrument at the 6.5-m MMT, which effectively increased the intermediate-redshift coverage out to $z=0.8$. DEEP2/DEEP3 provide a small but important extension of the galaxy sample to higher redshifts $z<1.1$ by several thousand spectra.

\section{Spectroscopic data analysis}

\paragraph{Post-processing of survey data}
For each spectral survey we created a post-processing routine to perform a conversion of spectra into a Virtual Observatory compatible format; relative flux calibration (everywhere except SDSS/eBOSS) using spectral sensitivity curves, which we constructed using either spectrophometric standard star observations or well-calibrated spectra from SDSS/eBOSS for a subset of overlapping sources; absolute flux calibration using broad-band flux measurements in the matching apertures. For the FAST spectra, we computed the sensitivity curves from observations of spectrophotometric standards available in the archive.

Spectra originating from FAST, DEEP2/DEEP3, 6dFGS, and 2dFGRS required a correction for telluric absorption. We used our own telluric correction procedure briefly described in \citet{2018MNRAS.477.4856A}: it fits an observed spectrum against a linear combination of synthetic stellar and galaxy spectra multiplied by a pre-computed grid of atmospheric transmission models from ESO {\sc skycalc} \citep{skycalc1} in the airmass-PWV (precipitable water vapor) parameter space

\paragraph{Full spectrum fitting}

The {\sc NBursts} \citep{CPSK07,2007MNRAS.376.1033C} method is a full spectrum fitting technique for simultaneous determination of the parameters of the stellar population and its kinematics from a spectrum of the galaxy integrated along the line of sight. This method uses a non-linear $\chi^2$ minimization for a spectrum and a model represented by a linear combination of one or more stellar population models and returns parameters of a line-of-sight velocity distribution (LOSVD) and a star formation history (SFH). The properties of emission lines (fluxes and widths/shapes) are fitted simultaneously with stellar populations, we can also fit multi-component emission lines, e.g. to model broad lines in active galactic nuclei. In RCSEDv2 we use several grids of stellar population models: simple stellar populations (SSPs) and exponentially declining SFHs from {\sc pegase.hr} \citep{2004A&A...425..881L}, SSPs from {\sc miles} \citep{2015MNRAS.449.1177V} with a resolution on the [Mg/Fe] enhancement, SSPs from {\sc e-miles} \citep{2016MNRAS.463.3409V} in a extended wavelength range. For small sub-samples of dwarf galaxies in clusters we also used self-consistent models of chemical evolution by \citet{2019arXiv190913460G}. For each spectroscopic survey, we matched the spectral resolution of the model grids to that of the survey data, which is crucial for unbiased estimates of the stellar velocity dispersion. Because of spectral resolving power varying from $R=800$ (2dFGRS) to $R=6500$ (DEEP2), the uncertainties of velocity dispersion measurements vary from one survey to another but typically stay under 5\%\ consistent with theoretical predictions \citep{2020PASP..132f4503C}. As an initial guess for stellar age, the most critical of the input parameters, we use an estimate based on the $g-r$ color of a galaxy.

\begin{figure}%[ht]
\vskip -5mm
\centering
\includegraphics[width=0.8\linewidth]{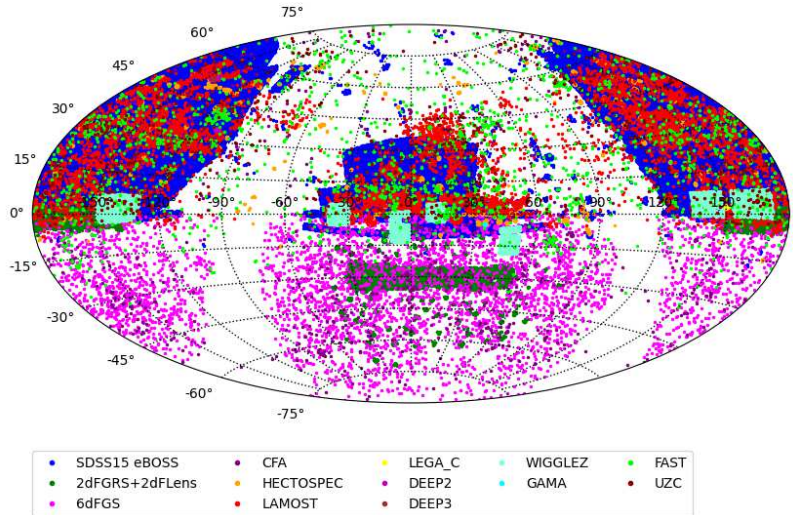}
\vskip -3mm
\caption{Sky coverage by the spectroscopic surveys included in RCSEDv2.
\label{X3-007_sky_rcsed}}
\end{figure}

\begin{figure}[ht!]
\begin{tabular}{rl}
\includegraphics[height=1\hsize, angle=270]{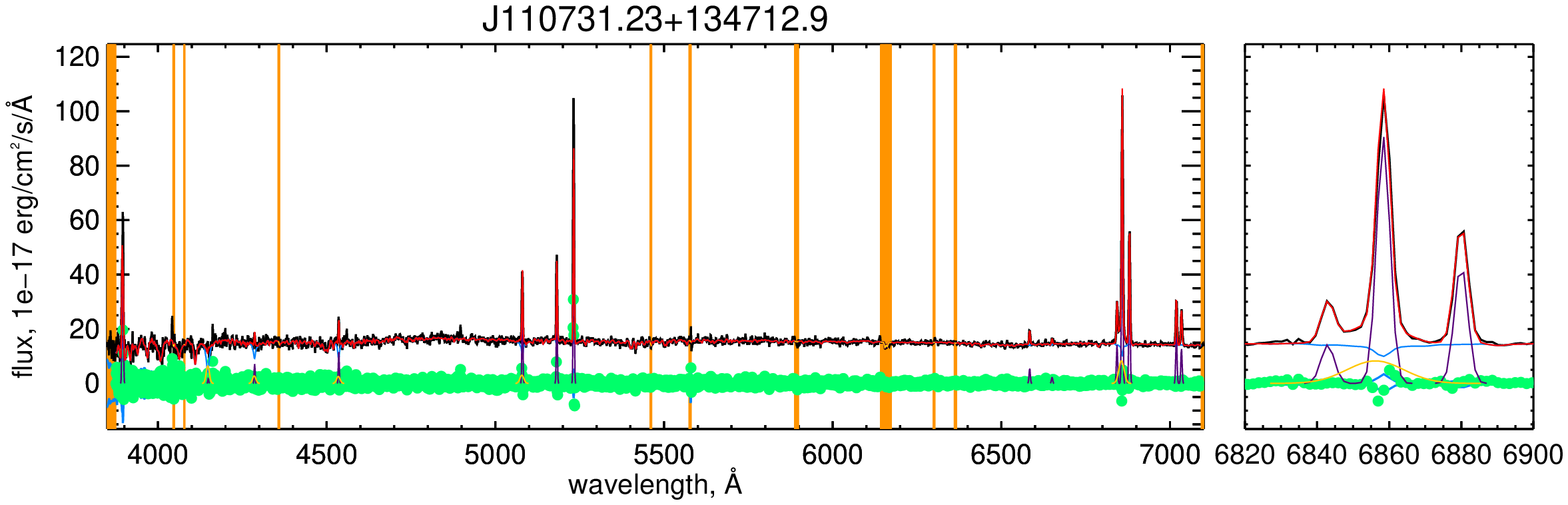} & \vspace{-0.55cm}
\\
\includegraphics[height=1\hsize, angle=270]{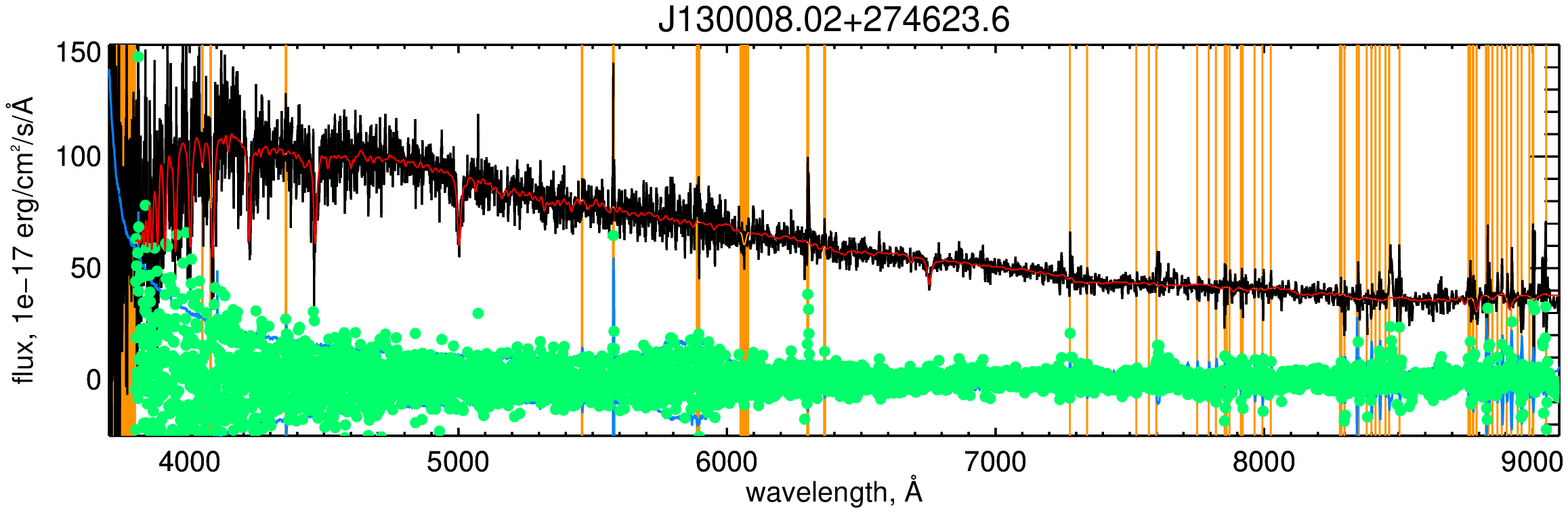} & \vspace{-0.55cm}\\
\includegraphics[height=1\hsize, angle=270]{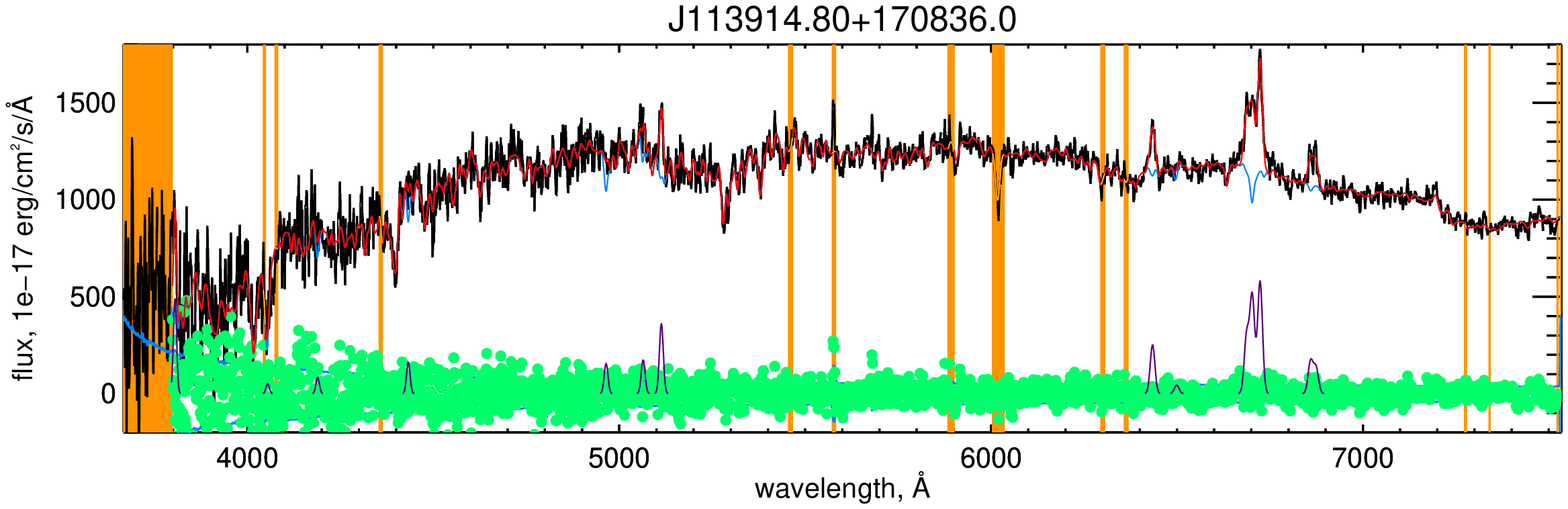} & 
\end{tabular}
{\caption{Examples of the full spectrum fitting using {\sc NBursts} for: (top) the low-mass galaxy hosting an active intermediate-mass black hole  J110731.23+134712.9 \citep{Chilingarian+18} from SDSS with a two-component emission-line spectrum; (middle) the extended low-mass post-starburst galaxy J130008.06+274623.8 in the Coma cluster \citep{2021NatAs.tmp..208G} from LAMOST; and (bottom) the giant low-surface brightness galaxy with an active galactic nucleus J113914.86+170837.0 \citep{2021MNRAS.503..830S} from FAST. The black, red, and green lines show a post-processed spectrum, its best-fitting models, and fitting residuals correspondingly.\label{X3-007_spec}}}
\end{figure}

\section{Galaxy properties derived from spectra}
Having run the full spectrum fitting, we compiled catalogs of galaxy properties derived from spectra and linked them to the main galaxy catalog in the RCSEDv2 database. The first group includes radial velocities, velocity dispersions, stellar ages and metallicities obtained from the fitting of each spectrum by 4 different grids of stellar population models: (i) {\sc miles} with [Mg/Fe] enhancements, (ii) {\sc e-miles}, (iii) {\sc pegase.hr} SSPs, and (iv) {\sc pegase.hr} exponentially declining SFHs (here we provide the exponential timescale instead of the stellar age). Emission lines were detected in about a third of all spectra. The second group of catalogs includes: emission-line fluxes, radial velocities / widths of emission lines, gas-phase metallicities and ionization parameters. All this information is accessible via the RCSEDv2 web-site and also through IVOA SSAP and TAP services for the post-processed spectra and properties derived from them respectively.

\acknowledgements This project is supported by the Russian Science Foundation Grant 19-12-00281 and the Interdisciplinary Scientific and Educational School of Moscow University ``Fundamental and Applied Space Research''. VG is grateful to the ADASS-XXXI organizing committee for providing financial aid to support his attendance of the conference.

\bibliography{X3-007}  % For BibTex

\end{document}